\documentclass{ws-p8-50x6-00}

\def\be{\begin{equation}}
\def\ee{\end{equation}}
\def\bea{\begin{eqnarray}}
\def\eea{\end{eqnarray}}

\begin{document}

\title{QCD CRITICAL POINT: WHAT IT TAKES TO DISCOVER}

\author{M.A. Stephanov}

\address{Institute for Theoretical Physics,
State University of New York,\\ 
Stony Brook, NY 11794-3840\\
E-mail: misha@insti.physics.sunysb.edu} 


\maketitle\abstracts{This report summarizes the results of the work done
in collaboration with K. Rajagopal and E. Shuryak.  We analyze the
physics behind the event-by-event fluctuations in heavy ion
collisions. Using thermodynamic description of the ensemble of events
we analyze and quantify various effects that are sensitive to the
proximity of the critical point on the phase diagram of QCD.  }

\section{Introduction}

The phase diagram of QCD in the temperature -- baryon chemical
potential plane has been a subject of intensified theoretical interest
recently. On the experimental front, with the advent of large
acceptance detectors such as NA49 and WA98 at CERN SPS, we are now able
to measure average event-by-event quantities which carry information
about thermodynamic properties of the system at freeze-out.
Our goal is to understand what we can learn 
about the phase diagram of QCD from this newly available and future
data.

\begin{figure}[hbt]
\psfig{figure=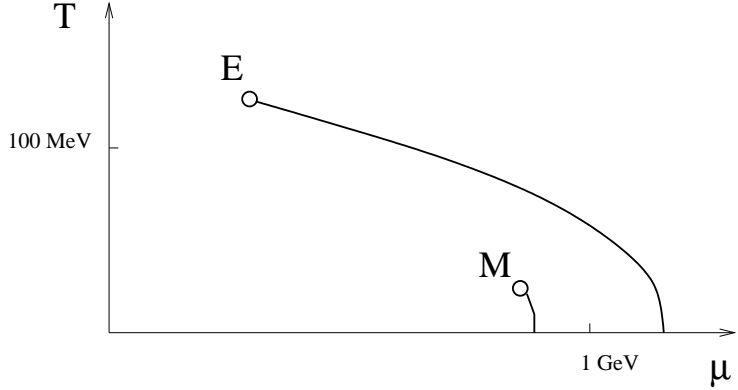,height=2in}
\caption[]{A schematic phase diagram of QCD~\cite{SRS}. Point E is the
critical end-point of the first order phase transition separating
quark-gluon and hadronic phases.\label{fig:pd}}
\end{figure}

The main focus of our analysis in \cite{SRS2} is on providing tools
for locating the critical point E on the phase diagram of QCD
(Fig.~\ref{fig:pd}) and studying its properties.  The possible
existence of such a point, as an endpoint of the first order
transition separating quark-gluon plasma from hadron matter, and its
universal critical properties have been pointed out recently in
\cite{BeRa97,HaJa97}.  In a previous letter, we have laid out the
basic ideas for finding this endpoint~\cite{SRS} in heavy ion
collision experiments.  The signatures proposed in~\cite{SRS} are
based on the fact that such a point is a genuine thermodynamic {\em
singularity} at which susceptibilities diverge and the order parameter
fluctuates on long wavelengths. The resulting signatures all share one
common property: they are {\em nonmonotonic\/} as a function of an
experimentally varied parameter such as the collision energy,
centrality, rapidity or ion size.

\section{Thermodynamic Fluctuations in an Ideal Bose Gas }
\label{sec:bose}

Most of our analysis is applied to the fluctuations of the observables
characterizing the multiplicity and momenta of the charged pions in
the final state of a heavy ion collision.
We begin building our tools by re-analyzing a text-book example of an
ideal Bose gas.  The basic fact is that every quantum state of
such a system  is completely
characterized by a set of occupation numbers, $n_p$. All observables
are functions of these numbers and thus all we need to know
is the fluctuations of $n_p$ from one member of the ensemble (one
event) to another:
\begin{equation}\label{dndn}
\langle n_p \rangle = {1\over e^{\epsilon_p/T} - 1},\quad
\langle \Delta n_p \Delta n_k\rangle 
=  \langle n_p\rangle (1 + \langle n_p\rangle) \delta_{pk}\equiv
{v}^2_p\delta_{pk}\ .
\end{equation}
The correlator $\langle \Delta n_p \Delta n_k\rangle$ is 
the central quantity which we calculate repeatedly, as we 
proceed beyond the ideal Bose gas approximation. 

The fluctuations of various extensive observables are given
in terms of the ``master correlator'' (\ref{dndn}):
\be\label{DeltaEfluctuation}
\langle (\Delta Q)^2\rangle = 
\sum_{pk} q_p q_k \langle \Delta n_p \Delta n_k\rangle =
\sum_p q_p^2 {v}^2_p 
\qquad\mbox{for}\qquad
Q = \sum_p q_p n_p\ .
\ee
More interestingly, the fluctuations of an intensive, or average, quantity,
such as energy or transverse momentum per particle, $q=Q/N$, are given by:
\begin{equation}\label{de}
\langle(\Delta q)^2\rangle 
= {1\over \langle N\rangle^2} 
\sum_p (q_p - \langle q\rangle)^2 {v}^2_p.
\end{equation}
Denoting by $\overline{ ^{\,}... _p }^{\rm inc}$ the average over
inclusive distribution $\langle n_p\rangle$ we see that
the ensemble average coincides with the inclusive one:
$\langle  q \rangle = \langle Q/N \rangle =
\overline{ q_p}^{\rm inc}$. 
This is not true for the fluctuation, however:
\begin{equation}\label{e=1p}
\langle(\Delta  q)^2\rangle= {1\over \langle N \rangle}
\overline{( q_p - \langle  q\rangle)^2 
(1+\langle n_p \rangle) }^{\rm inc}\ .
\end{equation}
We see that the event-by-event variance is larger than the suitably
rescaled (by $1/\langle N\rangle$) inclusive variance because of the
Bose enhancement factor $(1+\langle n_p\rangle)$. This order several
percent effect is very sensitive to the over-population of the pion
phase space characterized by the pion chemical potential $\mu_\pi$.

Another interesting quantity is the correlation between
the fluctuations of an average quantity and the total multiplicity $N$:
\begin{equation}\label{inclexclcorr}
\langle \Delta  q \Delta N\rangle =
\frac{1}{\langle N\rangle} \sum_p v_p^2 \left( q_p - 
\langle q\rangle\right)  = \frac{1}{\langle N\rangle} 
\sum_p \langle n_p\rangle^2 \left( q_p - 
\langle q\rangle\right)\ .
\end{equation}
Its value is totally due to the Bose effect, i.e., it would vanish in
the ideal classical gas limit $v^2_p=\langle n_p\rangle$.  For
example, for the ideal Bose gas of pions at a temperature $T=120$ MeV
the value of $\langle\Delta p_T \Delta N\rangle/ [\langle(\Delta
p_T)^2\rangle\langle(\Delta N)^2\rangle]^{1/2}$ is of the
order of a few percent and is negative.  In general, such
correlations, though small, are very sensitive to non-trivial effects,
such as Bose enhancement, as we have just seen, or to the effects
which we consider below such as as the energy conservation and thermal
contact or the interactions with the sigma field.

\section{Noncritical Thermodynamic Fluctuations\\ in Heavy Ion Collisions}

The next step in our characterization of the thermodynamic fluctuations
in heavy ion collisions is inclusion of pions from resonance
decays. The hadronic matter produced in a heavy ion collision
 is {\em not} simply an ideal
gas of pions. 
A number of approaches to heavy ion collisions have
successfully treated the matter at freeze-out as a
resonance gas in thermal equilibrium. The pions observed in the data
are then a sum of (i) ``direct pions'' which were
pions at freeze-out and (ii) ``resonance pions'' produced from
the decay of resonances after freeze-out.%

Our simulation of a resonance gas model \cite{SRS2} shows
that more than half of all observed pions come from resonance decays.
The resonances also have a dramatic effect on the
size of the multiplicity fluctuations. For an ideal classical gas
the ratio $(\Delta N)^2/N$ is equal to 1 and is only slightly
enhanced by the Bose effects. If some of the
pions are produced in bunches from resonances, which
themselves follow Poisson statistics, this ratio increases. We find:
\be\label{dN2/N}
\frac{\langle (\Delta N)^2\rangle}{\langle N\rangle} \approx 1.5 \ .
\ee

The experimental value from NA49 of this ratio  is 2.0.
This is much larger than the ideal gas value of 1.
The contribution of resonances is important to bring
this number up. However, there is still room for non-thermodynamic
fluctuations, such as fluctuations of impact parameter. Their
effect can be studied and separated by varying the centrality
cut using the zero degree calorimeter.

The $p_T$ spectrum of the resonance pions is close to the spectrum
of the direct ones. As a result resonance pions do not affect much
the shape of the spectrum, and in particular the width
of the inclusive distribution, which determines most of
the event-by-event fluctuation of the average $p_T$.
The resonances, however, dilute the Bose enhancement
effect by about a factor of two.

In order to compare the results with the experimental data one
has to take into account the effect of hydrodynamic flow.
This effect is not important for the multiplicity fluctuations.
However, it distorts the $p_T$ spectrum shifting it to larger $p_T$.
In the simplest approximation this can be treated as a ``blue shift''
of the spectrum. Essentially we assume that the effects of the flow
largely cancel in the ratio $v_{\rm inc} (p_T)/\langle p_T\rangle$.
This ratio in our simulation is equal 0.66.
The direct contribution from the
fluctuations of the flow velocity are small, order  2\% or so.
With the Bose enhancement included we obtain:
\begin{equation}
{v_{\rm inc} (p_T)\over\langle p_T\rangle} = 0.68.
\end{equation}
The experimental value obtained from NA49 data is 0.75.
We see that the major part of the observed fluctuation in $p_T$ 
is accounted for by the thermodynamic fluctuations. 
A large potential source of the discrepancy
is the ``blue shift'' approximation we used. This
approximation can be improved on in the future study.

Another very
important feature in the data is 
the value of the ratio 
of the scaled  event-by-event variation to the variance of the 
inclusive distribution:
\begin{equation}
F = {\langle N\rangle v^2_{\rm ebe}(p_T)\over v^2_{\rm inc}(p_T)} 
= 1.004 \pm 0.004.
\label{F}
\end{equation}
This is a remarkable fact, since the contribution of the Bose
enhancement (see Section~\ref{sec:bose}) to this ratio
is almost an order of magnitude larger than the statistical uncertainty.
Some mechanism must compensate for the Bose
enhancement. In the next section we find
a possible origin of this effect:
anti-correlations due to energy conservation and
thermal contact between the observed
pions and the rest of the system at freeze-out.

\section{Thermal Contact and Energy Conservation}
\label{contact}

In this Section,
we take a first step towards understanding
how the physics characteristic of the vicinity of the
critical point affects the event-by-event fluctuations.
Along the way, we quantify the effects of energy conservation
on the $p_T$-fluctuations.

We call the gas of direct pions ``system B'' and the rest of the system,
which includes  the
neutral pions, the resonances, the pions not in the experimental
acceptance and, if the freeze-out occurs near critical point, the
order parameter or sigma field, 
 --- ``system A''.
We observe system B, which is our ``thermometer''. 
The thermal contact of B with A and energy conservation
affects the ``master correlator'' (\ref{dndn}).
For example, $n_p$ cannot fluctuate completely independently
if the heat capacity $C_A$ of the system A is small. There is a constraint
on the total energy $E_B=\sum_p\epsilon_p n_p$ which gets stronger at
small $C_A$. The result for the master correlator we find is:
\begin{equation}\label{dndncorr}
\langle \Delta n_p \Delta n_k\rangle = {v}^2_p \delta_{pk}
- 
{{v}^2_p \epsilon_p {v}^2_k \epsilon_k \over 
T^2(C_A + C_B)}\ .
\end{equation}

Using this expression of the correlator we can now calculate the
effect of thermal contact and energy conservation on fluctuations
of various observables, such as mean $p_T$, for example.
In particular, we find that the anti-correlation introduced by this
effect reduces the value of the ratio $F$ defined in (\ref{F})
by: 
\begin{equation}
\Delta F_T \approx - {0.12\over C_A/C_B + 1} \ .
\label{DeltaF_T}
\end{equation}
If we take $C_A/C_B\sim3$ for orientation,
we find $\Delta F_T$ of the order
of $-3\%$, before taking into account the dilution by non-direct
pions. This effect is comparable in magnitude to the Bose enhancement,
and acts in the opposite direction.

This effect can be distinguished from other
effects, e.g., finite two-track resolution, also countering
the Bose enhancement, by the specific form of the
microscopic correlator (\ref{dndncorr}).  The effect of energy
conservation and thermal contact introduces an {\em off-diagonal} (in
$pk$ space, and also in isospin space)
anti-correlation. Some
amount of such anti-correlation is indeed observed in the NA49
data.

Another important point of (\ref{dndncorr}) is that as the freeze-out
approaches the critical point and $C_A$ becomes very large the
anti-correlation due to energy conservation disappears.

\section{Pions Near the Critical Point: Interaction with the Sigma Field}

In this section, unlike the previous sections, we shall consider
the situation in which the freeze-out occurs very close to the
critical point. This point is characterized by large long-wavelength
fluctuations of the sigma field (chiral condensate). We must 
take into account the effect of the $G\sigma\pi\pi$ interaction
between the pions and such a fluctuating field. We do that by
calculating the contribution of this effect to the ``master
correlator''. We find:
\begin{equation}\label{dndnsigma}
\langle{\Delta n_p\Delta n_k}\rangle = {v}^2_p\delta_{pk}
+ {1\over m_\sigma^2}{G^2\over T}
{{v}^2_p{v}^2_k\over \omega_p \omega_k}.
\end{equation}
We see that exchange of soft sigma field leads to a dramatic
off-diagonal correlation, the size of which grows as we approach
the critical point and $m_\sigma$ decreases. This correlation
takes over the off-diagonal anti-correlation discussed in the
previous section.

To quantify the effect of this correlation we computed
the contribution to the ratio $F$ (\ref{F}) from (\ref{dndnsigma}).
We find:
\begin{equation}\label{FTresultmu0}
\Delta F_\sigma =  0.14 \left(G_{\rm freeze-out}\over 300\ {\rm MeV}\right)^2 
\left(\xi_{\rm freeze-out}\over 6\ {\rm fm}\right)^2 
\qquad{\rm for~} \mu_\pi = 0\ ,
\end{equation}
This effect, similarly to the Bose enhancement, is sensitive to
over-population of the pion phase space characterized by $\mu_\pi$ and
increases by a factor 2.5 for $\mu_\pi=60$ MeV.  We estimate the size
of the coupling $G$ to be around 300 MeV near point E, and the mass
$m_\sigma$, bound by finite size effects, to be less than 6 fm. The
effect (\ref{FTresultmu0}) can easily exceed the present statistical
uncertainty in the data (\ref{F}) by 1-2 orders of magnitude.

It is important to note that we have calculated the effect of
critical fluctuations on $F$ because this ratio is being measured
in experiments, such as NA49. This observable is not
optimized for detection of critical fluctuations. It is easy to
understand that observables which are more sensitive to
small $p_T$ than $F$, and/or observables which are sensitive to
{\em off-diagonal} correlations in $pk$ space would show even
larger effect as the critical point is approached.

\section{Pions From Sigma Decay}

Near the critical endpoint, 
the excitations (quasiparticles) of the sigma field
are nearly massless and
are therefore numerous.  Because the pions are massive
at the critical point, these $\sigma$'s cannot immediately
decay into two pions and persist as the system expands after
freeze-out when it occurs near the critical point.
During the expansion, the in-medium $\sigma$ mass rises
towards its vacuum value and eventually exceeds the 
two pion threshold.  At this point the $\sigma$'s decay quickly,
yielding a non-thermal population of soft pions.

We estimate the mean momentum of these soft pions 
to be around 0.6$m_\pi$ and the total number
to be of the order of the number of direct pions (i.e.,
they should constitute up to a third of total observed pions
near the critical point). The multiplicity fluctuations
of these pions: $\langle (\Delta N)^2\rangle/ \langle N \rangle\ 
\approx  2.7$, are significantly larger than that of the
rest of the pions (\ref{dN2/N}).

\section{Conclusions}

In summary, our understanding of the thermodynamics of
QCD will be greatly enhanced by the detailed
study of event-by-event fluctuations in heavy
ion collisions.  We have estimated the influence of a number
of different physical effects on
the master correlator $\langle \Delta n_p\Delta n_k \rangle$.
This is itself measurable, but we have in addition used it
to make predictions for 
the fluctuations of observables which have been
measured at present, such as
$\langle (\Delta p_T)^2\rangle$ and $\langle (\Delta N)^2\rangle$
and also for the cross correlation $\langle \Delta N\Delta p_T\rangle$.

The signatures we analyze allow experiments
to map out distinctive features of the QCD phase diagram.
The striking example which we have considered in
detail is the effect of a second order critical end point.
The nonmonotonic
appearance and then disappearance of any one of the signatures 
of the critical fluctuations which we have described 
would be strong evidence for the critical point. 
Simultaneous detection of the
effects of the critical fluctuations on different observables
would turn strong evidence into a decisive discovery.



\begin{thebibliography}{99}
\bibitem{SRS2} M. Stephanov, K. Rajagopal, E. Shuryak,
		hep-ph/9903292.
\bibitem{BeRa97} J. Berges and K. Rajagopal, Nucl. Phys. {\bf B538} (1999)
215.

\bibitem{HaJa97} M. A. Halasz, A. D. Jackson, R. E. Shrock, M. A. Stephanov
and J. J. M. Verbaarschot, Phys. Rev. {\bf D58} (1998) 096007.


\bibitem{SRS} M. Stephanov, K. Rajagopal, E. Shuryak,
Phys. Rev. Lett. {\bf 81} (1998) 4816.
\end{thebibliography}
\end{document}